\documentclass{elsart3p-mod}

\usepackage{graphics}
\usepackage{graphicx}
\usepackage{amssymb}
\usepackage{upgreek}

\newcommand{\comment}[1]{}

\begin{document}

\begin{frontmatter}



\title{Deep-Sea Acoustic Neutrino Detection and the 
AMADEUS System as a Multi-Purpose Acoustic Array}


\author{Robert Lahmann\corauthref{rl} on behalf of the ANTARES Collaboration}
\corauth[rl]{robert.lahmann@physik.uni-erlangen.de}
%\ead{robert.lahmann@physik.uni-erlangen.de}
\address{Erlangen Centre for Astroparticle Physics (ECAP), Erwin-Rommel-Str. 1, 91058 Erlangen, GERMANY}

\begin{abstract}

The use of conventional neutrino telescope methods and technology for
detecting neutrinos with energies above 1\,EeV from astrophysical
sources would be prohibitively expensive and may turn out to be
technically not feasible. Acoustic detection is a promising alternative for
future deep-sea neutrino telescopes operating in this energy
regime. It utilises the effect that the energy deposit of the particle cascade evolving from a neutrino
interaction in water generates a coherently emitted sound wave with
frequency components in the range between about 1 and 50 kHz. The AMADEUS
(Antares Modules for Acoustic DEtection Under the Sea) project is
integrated into the ANTARES neutrino telescope and aims at the
investigation of techniques for acoustic particle detection in sea
water. The acoustic sensors of AMADEUS are using piezo elements and
are recording a broad-band signal with frequencies ranging up to 125
kHz. After an introduction to acoustic neutrino detection it will be
shown how an acoustic array similar to AMADEUS can be used for
positioning as well as acoustic particle
detection. Experience from AMADEUS and possibilities for a future
large scale neutrino telescope in the Mediterranean Sea will be
discussed.
\end{abstract}

\end{frontmatter}

% main text
\section{Deep-Sea Acoustic Neutrino Detection}
%\label{}
\subsection{Motivation}

The IceCube detector~\cite{bib:IceCube} currently under construction at the South Pole and the
KM3NeT Research Infrastructure~\cite{bib:KM3Net} planned in the Mediterranean 
Sea will push
the active target volume for neutrino detection towards the km$^3$-scale.
These detectors employ the well-established technique of detecting 
Cherenkov light from charged particle trajectories. % with photomultipliers.

While no high energy neutrinos from astrophysical sources have yet been 
identified, predictions 
about their flux can be deduced
from theoretical models as well as from the observed cosmic ray spectrum.

Several models predict a neutrino flux  
above $E\gtrsim 10^{18}$\,eV. Most prominently, 
Greisen~\cite{CR_Greisen}, Zatsepin and Kuzmin~\cite{CR_Zatsepin} (GZK)
predicted that cosmic ray protons interact with the cosmic microwave 
background,
which would lead to a cutoff in the observed cosmic ray energy spectrum at
values beyond $10^{18}$\,eV.
The Auger Collaboration has recently published results that are
consistent with this GZK cutoff~\cite{bib:Auger2008}. 

The consequence of this effect is a neutrino flux which for 
a cubic kilometre neutrino telescope would be at the edge of detectability.
Hence for the small fluxes of 
neutrinos with $E \gtrsim 10^{18}$\,eV  
new approaches are required.
One such approach is acoustic detection, which will be discussed in 
this paper.

\subsection{The Thermo-Acoustic Model}
The production of 
pressure waves by fast particles passing through liquids was 
predicted as early as 1957~\cite{Askariyan1}, 
leading to the development of the so-called
\emph{thermo-acoustic model} in the 
1970s~\cite{Askariyan2,Learned}. 
Around the same time, the effect was investigated experimentally
with proton pulses in fluid media~\cite{Sulak}.
According to the model, the energy deposition of charged
particles traversing liquids leads to a local heating of the medium
which can be regarded as instantaneous with respect to the
typical time scale of the acoustic signals. 
Because of the temperature change the medium
expands or contracts according to its volume expansion coefficient
$\alpha$. 
The accelerated motion of the heated medium produces a pressure
pulse which propagates through the medium.
The wave equation describing the pulse is \cite{Askariyan2}
\begin{equation}
  {\vec{\nabla}}^2 p(\vec{r},t) - \frac{1}{c_s^2} \cdot \frac{\partial^2 p(\vec{r},t)}{\partial t^2} =-
  \frac{\alpha}{C_p} \cdot \frac{\partial^2 \epsilon(\vec{r},t)}{\partial t^2}
\label{waveequation}
\end{equation}
and can be solved using the Kirchhoff 
integral~\cite{landau_tp,bib:niess_bertin2005}. 
Here $p(\vec{r},t)$ denotes the pressure at a given place
and time, $c_s$ the speed of sound in the medium, $C_p$ its specific
heat capacity and $\epsilon(\vec{r},t)$ the energy deposition density
of the particles. 
Attenuation of the signal during propagation can
be introduced by substituting ${\vec{\nabla}}^2 p$ by
${\vec{\nabla}}^2 \left(p+\frac{1}{\omega_0}\cdot\frac{\partial
    p}{\partial t}\right)$, with a characteristic attenuation
frequency $\omega_0$ in the GHz range~\cite{Learned}. 
The attenuation length
decreases with frequency and is on the order of 5\,km (1\,km) for 10\,kHz
(20\,kHz) signals.

The resulting pressure field is determined by the spatial and temporal
distribution of $\epsilon$ and by $c_s$, $C_p$ and $\alpha$.
The parameters $c_s$, $C_p$ and $\alpha$
exhibit a substantial temperature dependence. 
Laboratory investigations of the thermo-acoustic effect with particle and
laser beams
in different liquids have been documented by several 
authors; a list of references can be found in~\cite{bib:Graf_PhD_2008}.
The unique feature of the thermo-acoustic sound
generation -- the disappearance of the signal at $4^{\circ}\mathrm{C}$ in water,
due to the vanishing $\alpha$ at this
temperature -- was verified in~\cite{bib:graf_diplom}.

\begin{figure}[ht]
\centering
\includegraphics[width=7.6cm]{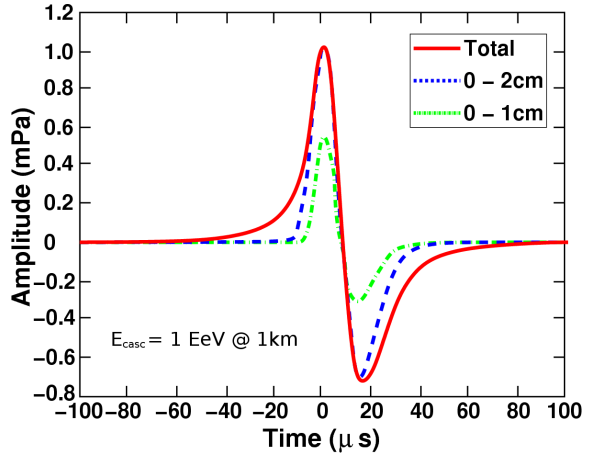}
\caption{
Simulated bipolar pressure pulse from a $10^{18}$\,eV cascade at
a radial distance of 1\,km.
Figure adapted from~\cite{bib:Acorne2007}.
}
\label{fig:bipolar_pulse}
\end{figure}

The spatial and temporal
distribution of $\epsilon$
is not accessible to laboratory experiments at the relevant 
energies $E \gtrsim 10^{18}$\,GeV 
and hence is subject to uncertainties. 
Simulations depend on both the extrapolation of parametrisations for lower
energies into this regime and on the transcription of simulations for
extended air showers to showers in water. Fig.~\ref{fig:bipolar_pulse} shows a bipolar
pressure pulse from a recent Monte Carlo simulation of neutrino interactions
in water for a cascade energy of 1\,EeV~\cite{bib:Acorne2007}. 
The amplitude was calculated at a distance of 1\,km
from the cascade in a plane
perpendicular to the shower axis at the shower maximum.

The cascades extend over a length of a few metres with a radius of only 
a few cm. In radial direction, i.e. perpendicular to the shower axis, 
the coherent superposition of the emitted sound
waves leads to the propagation of the sound within 
a flat disk (often referred to as {\em pancake}) with a 
small opening angle.
Assuming a linear dependence on the energy, the pressure pulse 
amplitude scales as 
\begin{equation}
p(\mathrm{1\,km}) \approx \frac{E_{casc}}{1\,\mathrm{EeV}}\,\mathrm{mPa} 
\end{equation}
with the energy spectral density peaking around 10\,kHz.

The simulation (Fig.~\ref{fig:bipolar_pulse}) shows that
roughly half of the pressure pulse is produced within a radius of 1\,cm
(dot-pointed line) of the cascade, whereas
the energy distribution within a radius of 2\,cm (broken line) is nearly 
completely responsible for the final signal shape (solid line).

\subsection{Acoustic Detection}

Two major advantages over an optical neutrino telescope
make acoustic detection worth studying. First,
the attenuation length is approximately one order of magnitude
higher for acoustic signals of cascades than for the Cherenkov light
(order of 1\,km to order of 100\,m) when comparing the relevant
frequency bands of the emissions. The second advantage is the
much simpler sensor design and readout electronics required for acoustic
measurements:  
No high voltage is required in the case of acoustic measurements
and time scales are in the 
$\upmu$s range for acoustics as compared to the ns range for optics. This
allows for online implementation of advanced signal processing
techniques. Efficient data filters are essential, as the signal
amplitude is relatively small compared to the acoustic background in
the sea, which complicates the unambiguous determination of the
signal.

Intensive studies are currently performed at various places to explore the
potential of the acoustic detection technique. 
In a recent survey~\cite{bib:Thompson2008}, 
an overview of these experimental acoustic
activities is given; 
for the remainder of this paper, the acoustic detection
system in ANTARES (AMADEUS) will be discussed.

\section{The Acoustic Detection System of ANTARES (AMADEUS)}
%\section{The Acoustic Setup}
\label{sec:acoustics}

\subsection{Goals}
\label{subsec:goals}
It is the declared main goal of the AMADEUS project to perform a 
feasibility study for a potential future large scale acoustic detector.
To this end, the following aims will be pursued:
\begin{itemize}
\item
Long term background investigations 
(rate of neutrino-like signals, localisation of sources);
\item
Investigation of signal correlations on different length scales;
\item
Development and tests of filter and reconstruction algorithms;
\item
Tests of different hydrophones and sensing methods;
\item
Studies of hybrid detection methods.
\end{itemize}
These goals were driving the design of the AMADEUS system, which
will now be discussed.

\subsection{AMADEUS as Part of ANTARES}
\label{sec:amadeus_part_of_antares}

\begin{figure}[ht]
\centering
\includegraphics[width=7.6cm]{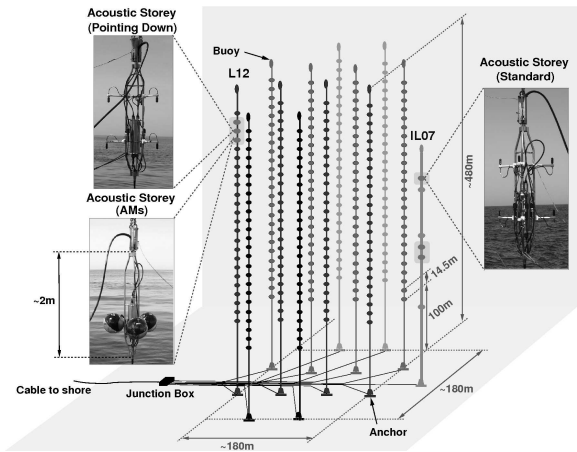}
\caption{A sketch of the ANTARES detector. 
The six Acoustic Storeys are highlighted and their three different setups
are shown. The setups will be described in Sec.~\ref{sec:acousensors}.}
\label{fig:ANTARES_schematic_all_storeys}
\end{figure}

AMADEUS is a part of the ANTARES neutrino telescope~\cite{bib:ANTARES} 
in the Mediterranean Sea, which was completed on May 30th, 2008. 
A sketch of the complete detector is shown
in Fig.~\ref{fig:ANTARES_schematic_all_storeys}.
ANTARES is located offshore, about 40\,km south of Toulon on the
French coast, at a depth of about 2500\,m. It comprises
12 vertical structures, the {\em detection lines} (or lines for short) 
plus a 13th line, called {\em Instrumentation Line} or {\em IL07},
which is equipped with instruments for monitoring the environment.
Each detection line holds
25 {\em storeys} that are arranged at equal distances of 14.5\,m along the 
line, interlinked by electro-mechanical-optical cables.
A standard storey consists of a titanium support structure, 
holding three {\em Optical Modules}, i.e. photomultiplier tubes (PMTs) 
inside water tight spheres, and one
{\em Local Control Module (LCM).} The LCM holds
the offshore electronics and the power supply within a cylindrical titanium container
(cf. Sec.~\ref{sec:offshore}). 
Each line is fixed on the sea floor by an anchor (called {\em Bottom String
Socket, BSS}) and held vertically by a buoy.

Acoustic sensing 
was integrated in form of {\em Acoustic Storeys} which are modified
versions of standard ANTARES storeys, replacing the PMTs by hydrophones and
using custom-designed electronics for the digitisation and preprocessing of 
the analogue signals. Details will be discussed in the following 
subsections.

The three Acoustic Storeys on the IL07 started data taking when the connection
to shore
of the IL07 was established on Dec. 5th, 2007, the Acoustic Storeys on 
Line 12 (L12) 
were connected to shore during the completion of ANTARES in May 2008.
AMADEUS is now fully functional with 34 of its 36 sensors working.

\subsection{Description of the System}

It has been a fundamental design principle of the AMADEUS system to
make use of standard ANTARES hard- and software as much as possible in order
to  minimise the effort for design and  engineering
and to reduce the failure risks for both ANTARES and AMADEUS
by keeping the need for additional quality assurance and control
measures to a minimum.
In order to integrate the system successfully into the ANTARES detector, 
design efforts in three basic areas was necessary: First,
the development of hydrophones that replace the PMTs
of standard ANTARES storeys; second, the
development of an offshore Acoustic ADC and preprocessing board; 
third, the development of on- and offline software. 
These subjects will be discussed in more detail in
Sections \ref{sec:acousensors},
\ref{sec:acouadc-board}, and \ref{sec:onshore},
respectively.

The final system has
full detection capabilities (such as time synchronisation and a %sophisticated, 
continuously operating data acquisition) and is scalable to a larger
number of Acoustic Storeys. 
It combines local clusters of acoustic sensors with large cluster spacings,
allowing for fast direction reconstruction with individual storeys that 
then can be combined to reconstruct the position of a source.

\begin{figure}[ht]
\centering
\includegraphics[height=7.2cm]{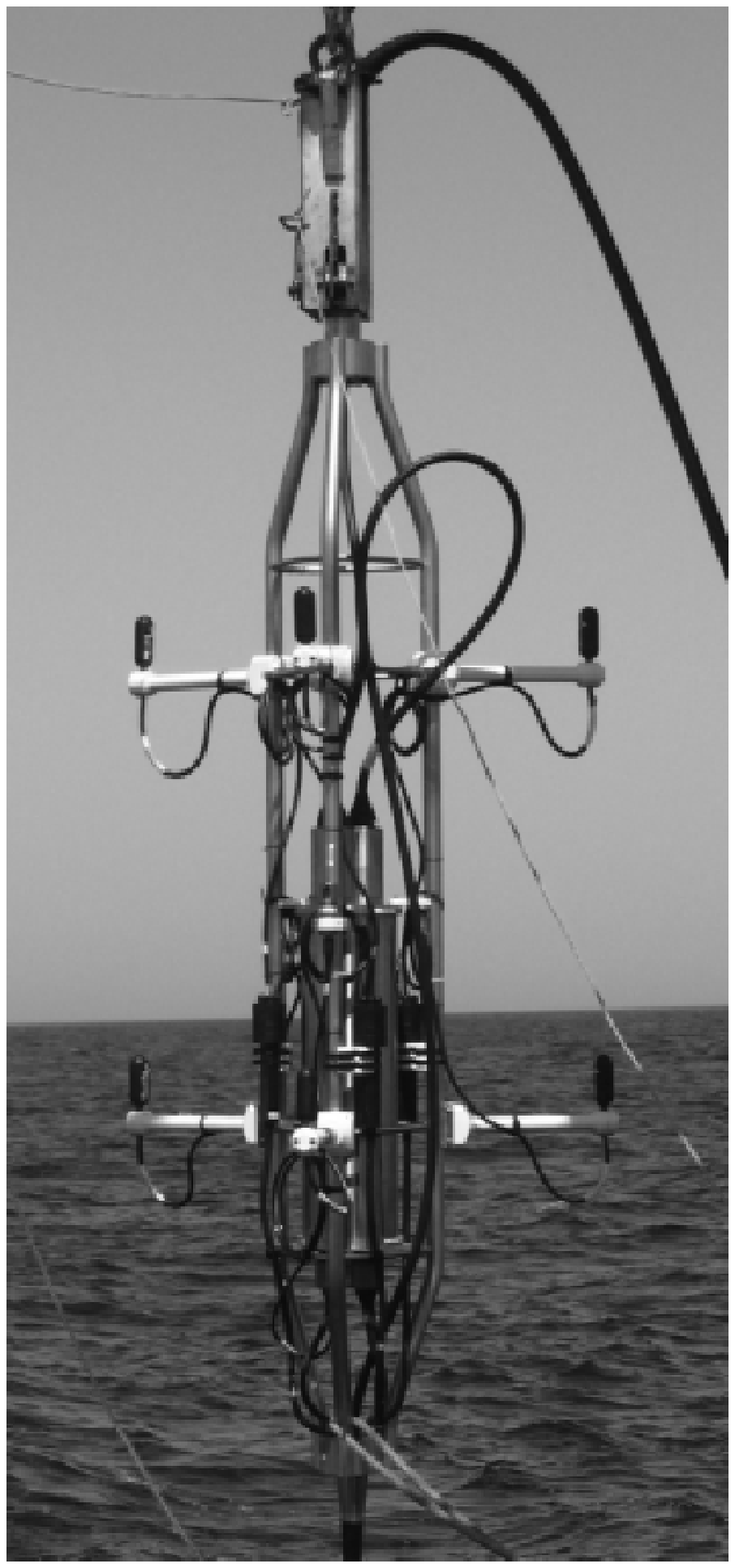}
\hspace{0.75mm}
\includegraphics[height=7.2cm]{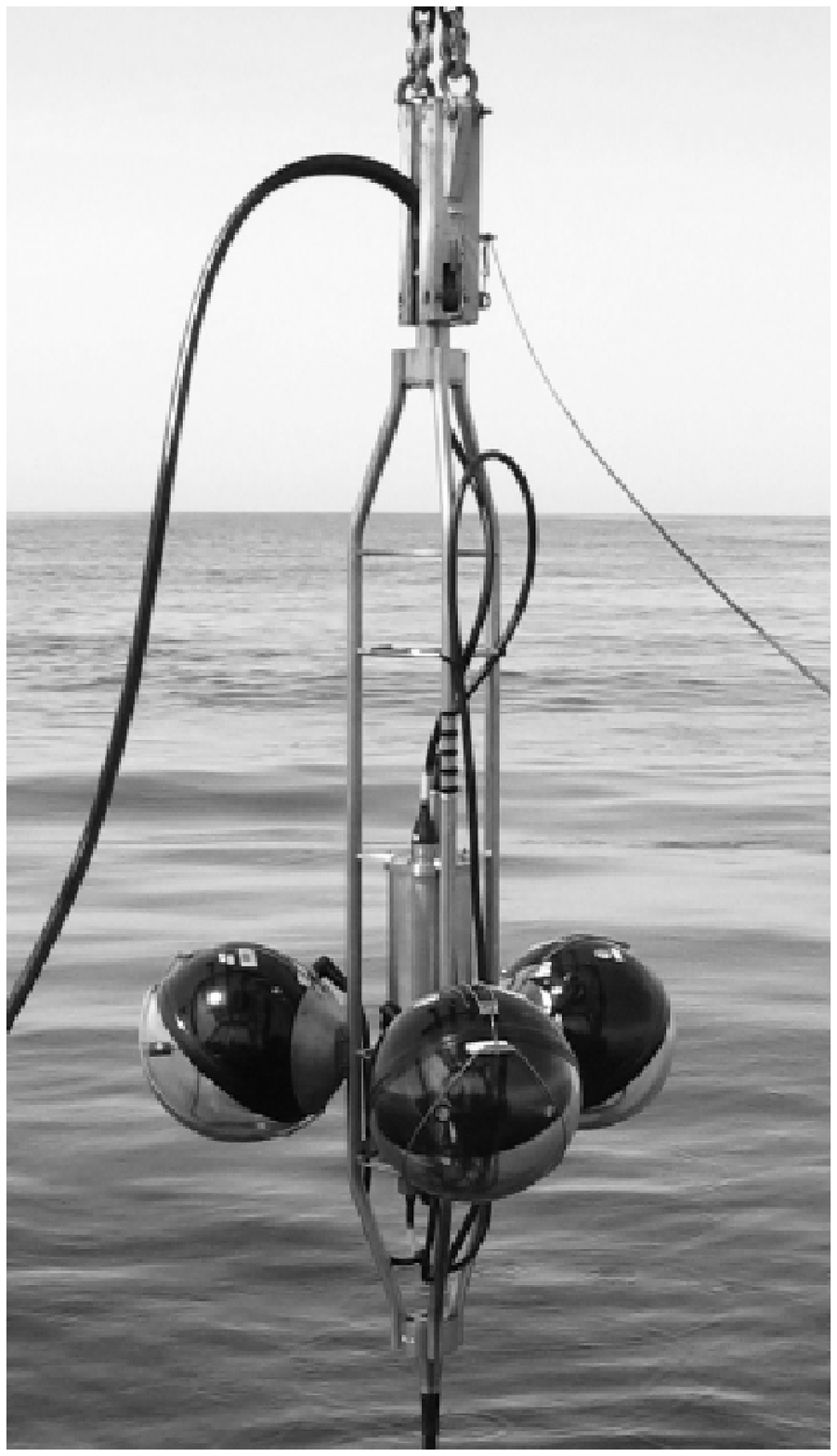}
\caption{Storey 3 of the IL07 (left) and Storey 21 of L12 (right) during their deployment.}
\label{fig:antares_storey_acou}
\end{figure}

Each Acoustic Storey is equipped with six acoustic sensors with interspacings
on the order of 1\,m.
The Acoustic Storeys 2, 3, and 6 on the IL07
are located at 180\,m, 195\,m, and 305\,m above the sea floor, respectively.
Line 12 is anchored at a horizontal distance of about 240\,m from the IL07,
with the Acoustic Storeys positioned at a heights of 380\,m, 395\,m, and 410\,m.
With this setup, the maximum distance between two Acoustic Storeys is 340\,m.
Two of the six Acoustic Storeys are shown in 
Fig.~\ref{fig:antares_storey_acou}.

\subsection{The Acoustic Sensors}
\label{sec:acousensors}

Two types of acoustic sensors are used in AMADEUS:
hydrophones and so-called {\em Acoustic
Modules} ({AMs}, cf. Fig. \ref{fig:antares_storey_acou}). In both cases, the
sensors are based    on piezo-electrical ceramics that convert pressure
waves into voltage signals, which are then amplified for 
readout~\cite{bib:hoessl2006}. The ceramics and amplifiers are coated in 
polymer plastics in the case of the hydrophones. For the AMs they are glued to
the inside of the same spheres used for the Optical Modules 
of ANTARES. The latter non-conventional design was inspired by the idea 
to investigate an option for acoustic sensing that can be combined 
with a PMT in the same housing. 

In order to obtain a complete and uniform 2$\pi$-coverage of the  azimuthal angle 
$\phi$, the 6 sensors are distributed over the 3 AMs of the storey 
within a plane perpendicular to the longitudinal axis of a storey at
angles of 60$^\circ$ in $\phi$.\footnote{
For reasons such as limited data rate and the pre-existing 
design of the ANTARES
offshore electronics, implementing more than 2 sensors per AM would have increased the
technical effort disproportionately.}

The three Acoustic Storeys on the IL07 house hydrophones only, whereas Storey
21 (counting from the bottom) of Line 12 holds AMs 
(cf. Fig.~\ref{fig:ANTARES_schematic_all_storeys}). 
In Storey 22 of Line 12, the hydrophones were mounted
with their cable junction, where the sensitivity is largely reduced,
pointing upwards. This allows for investigations of the directionality of background from ambient
noise, which is expected to come mainly from the sea surface.

Three of the five storeys holding hydrophones are equipped  with 
commercial hydrophones\footnote{
The hydrophones were produced by High Tech Inc (HTI) in Gulfport, MS (USA) 
according to specifications from Erlangen.}
and the other two with hydrophones developed and produced at the Erlangen 
Centre for Astroparticle Physics (ECAP).

\begin{figure}[tbh]
\centering
\includegraphics[height=6.3cm]{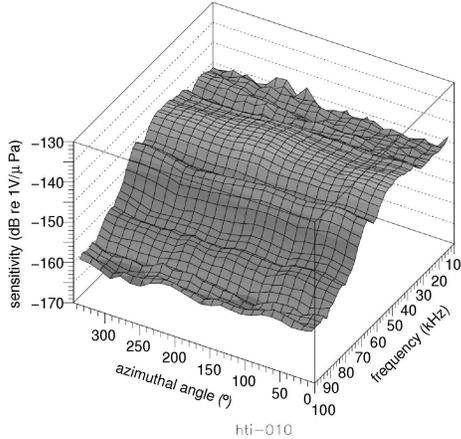}
\caption{
Sensitivity of a typical AMADEUS hydrophone 
as a function of frequency and azimuthal angle. 
}
\label{fig:hydrophone_sensitivity}
\end{figure}

All acoustic sensors are tuned to be sensitive over the whole
frequency range of interest from 1 to 50\,kHz with a typical sensitivity around
$-145$\,dB\,re.\,1V/$\upmu$Pa  (including preamplifier) and to have a low noise level
\cite{bib:naumann_arena05}. 
The sensitivity of a typical hydrophone 
is shown in Fig.~\ref{fig:hydrophone_sensitivity} as a function  of frequency
and the azimuthal angle~\cite{bib:naumann_phd}. 
For a given frequency, the distribution is essentially flat on a 3\,dB level.
\subsection{Offshore Electronics and Acoustic Data Acquisition}
\label{sec:offshore}

In the ANTARES DAQ scheme~\cite{bib:antares_daq},
 the digitisation is conducted within the 
offshore electronics container (LCM, cf. Sec.~\ref{sec:amadeus_part_of_antares})
on each storey by several custom-designed electronics boards, which
send all digitised data to shore where data reduction is performed and the
data is searched for events of interest.
With its capability
of timing resolutions on a nanosecond-scale\footnote{ 
The clock system is in fact capable of providing sub-nanosecond precision
for the synchronisation of the optical data recorded by the PMTs. This
precision, however, is not required for the acoustic data.}
and transmission
of several MByte per second and per storey, it is perfectly suited for the
acquisition of acoustic data. In addition, in each LCM 
a {\em Compass board} measures the tilt and 
the orientation of the storey by measuring the three components
of the Earth's magnetic field.

For the digitisation and preprocessing of the acoustic signals and 
for feeding them into the ANTARES data stream, 
the so-called {\it AcouADC-board} was designed.
Each board processes the differential signals from two acoustic 
sensors, which results in a total of three such boards
per storey.

\begin{figure}[ht]
\centering
\includegraphics[width=7.7cm]{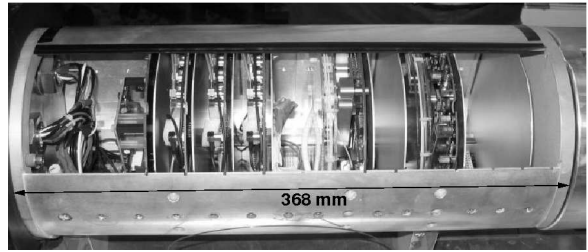}
\hspace{2pc}%
%\begin{minipage}[b]{8cm}
  \caption{
An LCM equipped with AcouADC-boards before insertion into its titanium housing. 
Visible on the left from the inside 
of the container are two of the three connectors which on the outside
are connected to two acoustic sensors each.
The boards are described in the text.
\label{fig_Acou_LCM}}
%\end{minipage}
\end{figure}

Figure~\ref{fig_Acou_LCM} shows the fully equipped LCM of an Acoustic Storey.
From the left to the right, the following boards are installed:
The Compass board;
3 AcouADC boards; a {\em Data Acquisition (DAQ) board} that sends the data to 
shore;
and a {\em Clock board} that provides the timing signals to correlate 
measurements performed in different storeys.

\subsection{The AcouADC-Board}
\label{sec:acouadc-board}
\begin{figure}[ht]
\centering
\includegraphics[width=5.5cm]{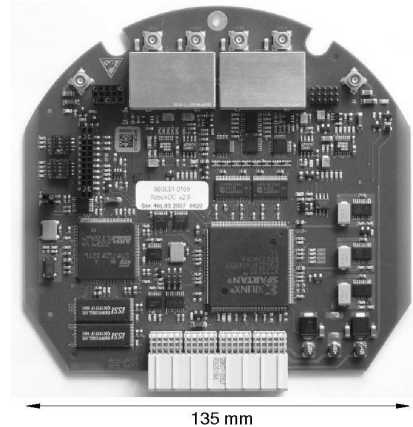}\hspace{2pc}%
%\begin{minipage}[b]{8cm}
  \caption{
  An AcouADC board. 
 %The width of the board is 135\,mm. 
\label{fig_AcouADC_board}}
%\end{minipage}
\end{figure}

The AcouADC-board is shown in Fig.~\ref{fig_AcouADC_board}. 
It consists of an analogue and a digital part. 
The
analogue part amplifies the voltage signals coming from the acoustic
sensors by adjustable factors between 1 and 562 and filters the
resulting signal.  
To protect the analogue part from potential electromagnetic interference, 
it is shielded by metal covers.
The system has low noise and is designed to be -- together with the 
sensors -- sensitive to the acoustic background of
the deep sea over a wide frequency-range.  %(approx. 1 to 100\,kHz).
The dynamic range achieved is from the
order of 1\,mPa to the order of 10\,Pa in RMS over the frequency range
from 1 to 100\,kHz. 

The analogue filter 
suppresses frequencies below approx. 4\,kHz and above approx.\
130\,kHz. The high-pass part cuts into the trailing edge of the low
frequency noise of the deep-sea acoustic background~\cite{urick} and
thus protects the system from saturation. The low-pass part
efficiently suppresses frequencies above the Nyquist frequency of
250\,kHz for the sampling rate of 500\,kSamples per second (kSPS). 
Within the passband, the filter response is essentially flat with a linear 
phase response.

Every individual component of the complete data taking chain was 
calibrated in the laboratory prior to deployment.
With the deduced transfer
function of the system it is possible to reconstruct 
the acoustic signal from the recorded one with high precision 
within the sensitive frequency range of the setup.

The digital part of the AcouADC-board digitises and processes the acoustic data. It is
designed to be highly flexible by employing a micro controller ($\upmu$C) 
and a field programmable gate array (FPGA) as data
processor. The $\upmu$C can be controlled with the onshore control
software and is used to adjust settings of the analogue part and the data
processing.  
Furthermore, the $\upmu$C can be used to update
the firmware of the FPGA in situ. 

The digitisation is done by two 16-bit ADCs for the two input channels.
The digitised data from the two channels is read out in parallel by the FPGA 
and further processed for transmission to the DAQ board.
The DAQ board then handles the transmission of the data to the 
onshore data processing servers.

In standard mode, the digitised data is down-sampled to 250\,kSPS by the 
FPGA, corresponding to a downsampling by a factor of 2. 
Hence the frequency spectrum of interest from 1 to 100\,kHz is fully contained in the data. 

Onshore a dedicated computer cluster is used to process and store
the acoustic data arriving from the storeys and to control the offshore
DAQ. This is discussed below.

\subsection{Onshore Data Processing}
\label{sec:onshore}
AMADEUS follows the same ``all data to shore'' strategy as ANTARES; 
the offshore data arrives via the TCP/IP protocol at a Gigabit switch 
in the ANTARES control room, where the acoustic data is separated from 
the standard ANTARES data and routed to the acoustic computer cluster.

The cluster currently consists of four servers of which two are used for 
data filtering. The filtering has the task of 
reducing the raw data rate of about 1.6~TB/day to about ~10 GB/day for 
storage. Currently, three filter schemes are implemented~\cite{Neff_diplom}:
A minimum bias trigger which records 10\,s of continuous data every 30\,min;
a threshold trigger; a cross correlation trigger, which searches for 
the expected bipolar signal of a neutrino. 
The filtering requires coincidences of several hydrophones on each storey and
can be extended to require coincidences between storeys on the same line. 
All parameters can be freely adjusted.
The concept of local clusters (i.e. the Acoustic Storeys) has proven very
efficient for fast (online) processing.
Furthermore, all components are scalable which makes 
the system extremely flexible. Additional servers
can be added or the existing ones can be replaced by the latest models
if more sophisticated filter algorithms are to be implemented.
In principle it is also possible to move parts of the filter into the FPGA
of the AcouADC board, thereby implementing an offshore trigger which reduces
the size of the data stream sent to shore.

Just like ANTARES, AMADEUS can be controlled via the Internet from 
essentially any place in the world and is currently operated from Erlangen. 
Data are centrally stored and are available remotely as well.

\subsection{First Results}
\subsubsection{Noise Measurements}
The ambient noise level in the frequency range from about 200\,Hz to 50\,kHz 
in the deep sea is assumed to be mainly determined
by the agitation of the sea surface, i.e. by waves, spray and 
precipitation~\cite{urick2}.
To verify these assumptions, the correlation 
between the weather and the RMS of the ambient noise recorded by AMADEUS
was investigated.
Weather data at the Hy\`eres airport at the French coast, about 30\,km north of 
the ANTARES site, is continuously logged by the AMADEUS onshore software.
New data is made available every hour.
%It is an important verification
%of the assumptions about the origin of the noise to investigate the correlation 
%between the weather and the RMS of the ambient noise recorded by AMADEUS.
%

\begin{figure}[ht]
\centering
\includegraphics[width=7.8cm]{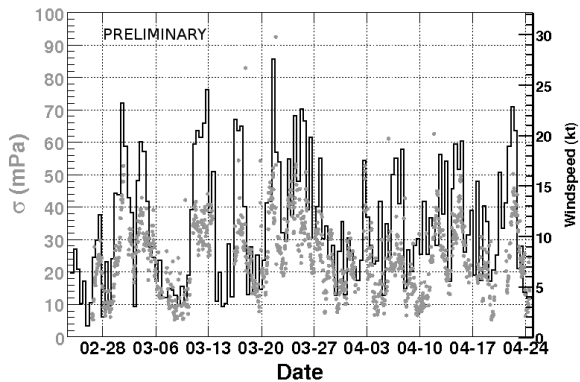}
\caption{RMS noise recorded by a 
characteristic hydrophone (light dots) and windspeed (dark solid line) as a function of time, 
showing the correlation between the two quantities.}
\label{fig:correlation_wind_rms_phd}
\end{figure}

An upgrade of the study performed in~\cite{bib:Graf_PhD_2008}
is shown in Fig.~\ref{fig:correlation_wind_rms_phd}.
Each point in the figure represents the RMS noise of a 10\,s sample of data, 
calculated by integrating the power spectral density (PSD) 
in the frequency range from 1 to 50\,kHz.
The correlation between the wind speed and the hydrophone noise is
found to be about 80\%.

The results are consistent with those reported for other deep-sea sites in
~\cite{SAUND2008, ONDE2008}.

\subsubsection{Positioning of Acoustic Storeys}
A further important measurement with the AMADEUS system is the accurate 
determination of the relative positions of the Acoustic Storeys within the 
ANTARES detector ({\em positioning}).
The ANTARES positioning system~\cite{bib:ardid_these_procs} uses 
transceivers (so-called pingers) at the BSS 
(cf. Sec.~\ref{sec:amadeus_part_of_antares}) of each
line in combination with 5 acoustic receivers (positioning hydrophones)
arranged along each detection line. 
The pingers emit tone bursts at 9 well-defined frequencies between 
44\,522\,Hz and 60\,235\,Hz which
can be used by AMADEUS to determine the positions of the Acoustic Storeys.

With 6 hydrophones, a complete reconstruction (position and three angles)
of each Acoustic Storey can be done using the pinger signals.
This is not only an interesting task on its own,
but is in fact important for an accurate positioning of unknown 
sources as will be described below.
The presence of the Compass boards (cf.~Sec.\ref{sec:offshore}) 
allows for cross checks of the results.

\begin{figure}[ht]
\centering
\includegraphics[width=7.7cm]{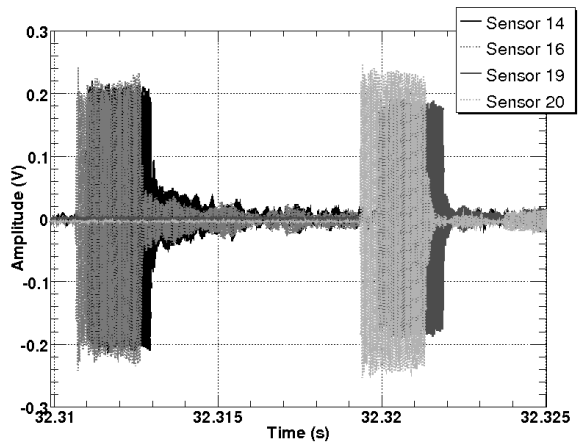}
\caption{Pinger signals received by sensors 19, 20 
(hydrophones in Storey 22 of Line 12), and sensors 14, 16 
(sensors of AMs in Storey 21 of Line 12). 
The sensors in the lower Storey 21
receive the signal earlier than those in Storey 22. Hydrophone 20 is
positioned in the bottom row, hydrophone 19 in the top row of the storey
(cf. Fig.~\ref{fig:antares_storey_acou}, left). 
Sensors 14 and 16 are located in
different AMs, separated by about 120$^\circ$ in $\phi$ (cf. Fig.~\ref{fig:antares_storey_acou}, right).  
\label{pinger_hit_am_full}
}
\end{figure}

Figure~\ref{pinger_hit_am_full} shows the signal emitted from one pinger as 
recorded by two sensors in the AMs of Storey 21 and 
by two hydrophones of Storey 22 of Line 12.
One can clearly observe the different arrival times of the signal 
(corresponding to different travel times of the sound from the pinger) 
between the storeys.
% (separated by a vertical distance of 14.5\,m). 
For the two exemplary sensors of one storey, the smaller differences
in arrival times are also clearly visible.

Work on two methods for the positioning is currently in progress: 
First, the differences between the absolute time of signal emission and reception from several pingers 
are used to reconstruct the position of each hydrophone individually.
Second, only the differences in arrival times of a pinger signal in the
6 hydrophones of a storey are used to reconstruct the direction of a pinger signal. 
Then the reconstructed directions are matched with the pinger pattern on the sea floor.
The second method employs the same algorithms 
used for position reconstruction of unknown sources; this will be described now.

\subsubsection{Position Reconstruction of Sources}

Position reconstruction of point sources will be done by first reconstructing their
direction from individual storeys and then combining the reconstructed directions
from three or more storeys.

To find the direction of point sources, a beam forming algorithm is 
used~\cite{richardt_diplom}. 
It is designed to reconstruct plane waves, which 
for the geometry of a storey is a reasonable assumption for sources with
distances $\gtrsim 100\,m$.

\begin{figure}[ht]
\centering
\includegraphics[width=7.5cm]{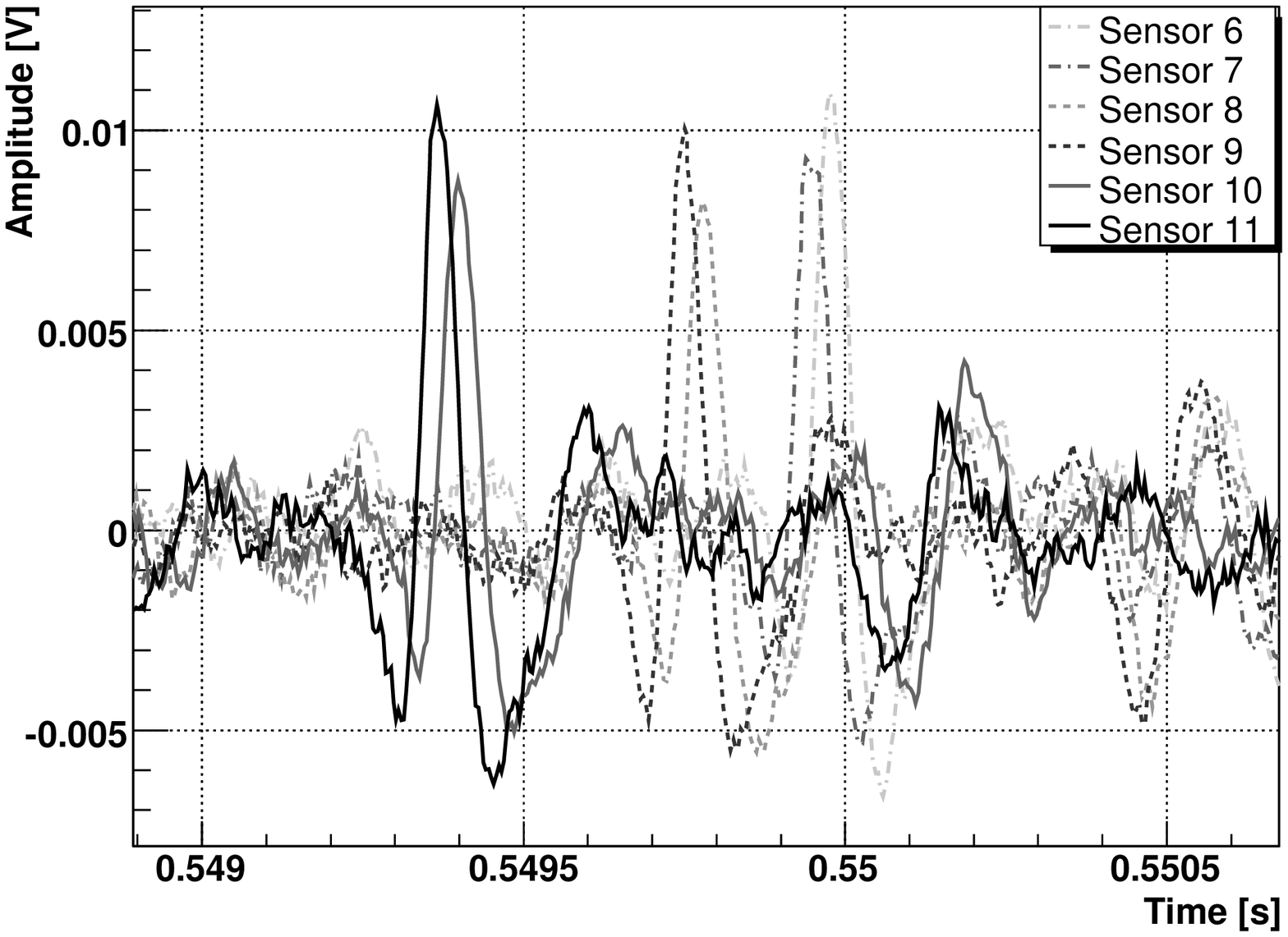}
\includegraphics[width=7.7cm]{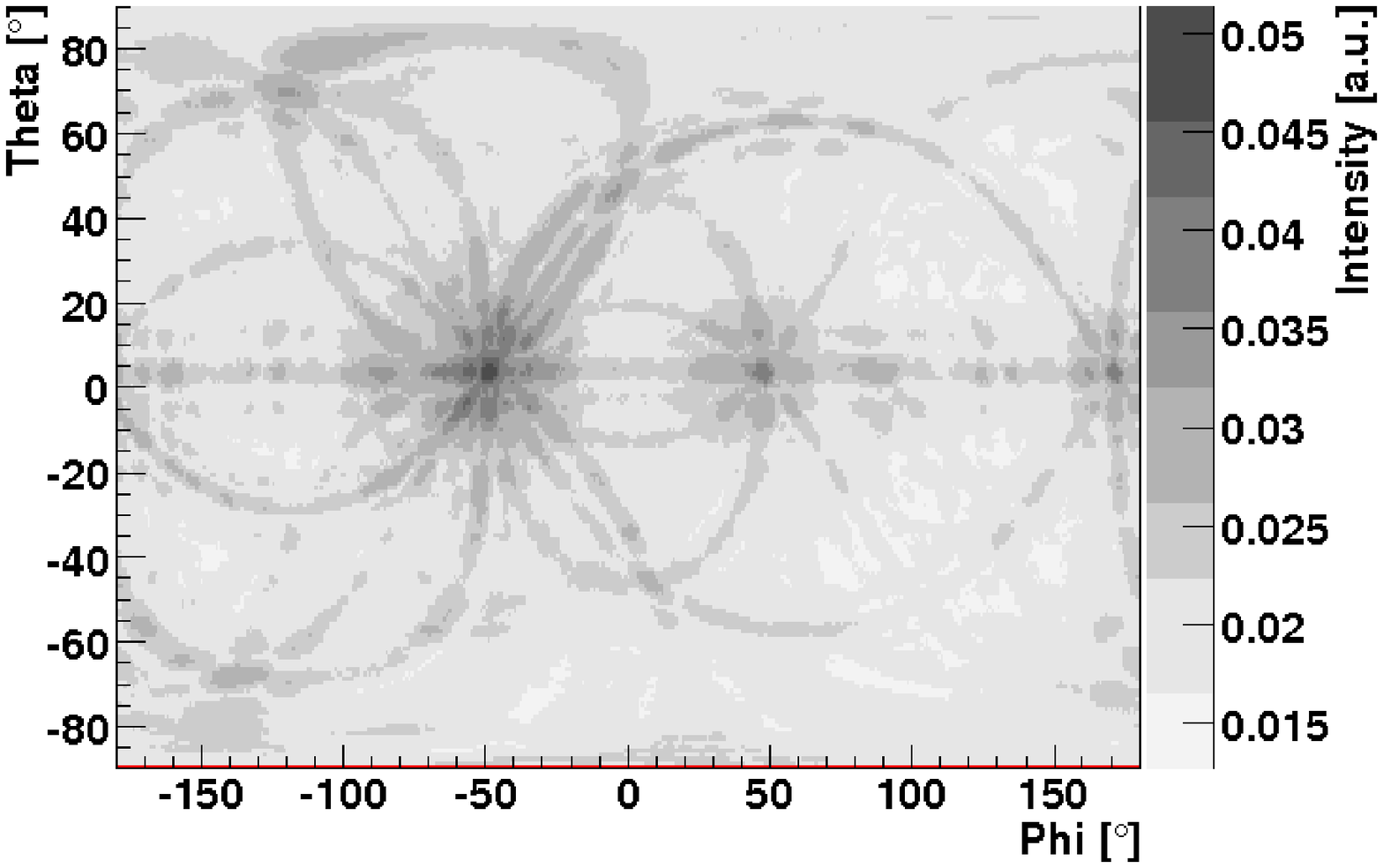}
\caption{
Reconstruction of the direction of an exemplary source. 
Top: Signals recorded with the hydrophones of the topmost storey of the IL07;
Bottom: Result of the beamforming algorithm.
}
\label{transient_signal_reco}
\end{figure}

The top plot of Fig.~\ref{transient_signal_reco} shows an exemplary signal as recorded by 
the topmost Acoustic Storey of the IL07. Sensors 6 and 7, 8 and 9, 10 and 11, respectively, 
are attached to
the same rod of the storey (cf. Fig.~\ref{fig:antares_storey_acou}); 
hydrophones with even numbers are located at the bottom.
Signals arrive at the hydrophones positioned at the same rod almost at the same time,
indicating that the direction of the source is close to horizontal.

The bottom plot shows the result of the beam forming algorithm. 
The intensity at a given
solid angle corresponds to 
the probability that the source is located in that direction. 
Bands of increased intensity come from potential source directions where 
the signals from only two or three hydrophones coincide.

The most probable direction, at $\theta \approx 5^\circ $ and 
$\phi \approx -50^\circ $ is consistent with the intuitive interpretation
of the top plot.

\section{Conclusions and Outlook}
The AMADEUS system, which is dedicated to the investigation of
acoustic neutrino detection techniques, has been successfully 
installed in the ANTARES detector.
Except for its small size, the system has all features required for
an acoustic neutrino telescope and hence is excellently suited for 
a feasibility study of a potential future large scale acoustic 
neutrino telescope.

AMADEUS can be used as a multi
purpose device for studies of neutrino detection techniques, 
position reconstruction, and marine research; for the latter, which was not discussed
in this paper, discussions with external partners with expertise in the field are ongoing.

First results were presented which demonstrate the potential of AMADEUS.
In particular, the Acoustic Modules allow for acoustic 
measurements without additional mechanical structures which might be an
option for KM3NeT.

\section{Acknowledgements}
This study was supported by the German government through BMBF grant 05CN5WE1/7.
The author wishes to thank the organizers of the VLVnT08 for a most interesting and
well organised workshop.

% \appendix

% \section{}
% \label{}


\begin{thebibliography}{00}
\raggedright
\bibitem{bib:IceCube}
IceCube Coll., A. Silvestri \etal, Mod. Phys. Lett. {\bf A22} (2007) 1769.
\bibitem{bib:KM3Net}
KM3NeT homepage, \verb+ http://www.km3net.org+.

\bibitem{CR_Greisen}
K.~Greisen, Phys. Rev. Lett. {\bf 16} (1966) 748.


\bibitem{CR_Zatsepin}
G.T.~Zatsepin and V.A.~Kuz'min, JETP Lett. {\bf 4} (1966) 78;
\newblock  \v{Z}. \`Eksp. Teor. Fiz., (1966) 114.


\bibitem{bib:Auger2008}
Auger Coll., A. Watson \etal.,
%``Recent results from the Pierre Auger Observatory: 
%Including comparisons with data from AGASA and HiRes'',
Nucl. Instr. Meth. A {\bf 588} (2008) 221. 


\bibitem{Askariyan1}
G.A. Askariyan,
%title="Hydrodynamic Radiation From the Tracks of Ionizing Particles in Stable Liquids",
Sov. J. At. En. {\bf 3} (1957) 921.

\bibitem{Askariyan2}
G.A. Askariyan, B.A. Dolgoshein \etal,
Nucl. Instr. Meth. {\bf 164} (1979) 267. 

\bibitem{Learned}
J.G.~Learned,
%title="Acoustic Radiation by Charged Atomic Particles in Liquids: An Analysis",
Phys. Rev. {\bf 19}, (1979) 3293.

\bibitem{Sulak}
L. Sulak, T. Armstrong \etal,
%title="Experimental Studies of the Acoustic Signature of Proton Beams Traversing fluid media",
Nucl. Instr. Meth. {\bf 161} (1979) 203. 

\bibitem{landau_tp}
L.D.~Landau and E.M.~Lifshitz,
  {\it Course of Theoretical Physics, Vol.~6: Fluid Mechanics},
 Pergamon Press, Oxford,
 1959.

\bibitem{bib:niess_bertin2005} 
V.~Niess and V.~Bertin,
%\newblock  {\em Underwater Acoustic Detection of Ultra High Energy Neutrinos},
  Astropart. Phys. {\bf 26}  (2006) 243.
%(preprint arXiv:astro-ph/0511617v3).

\bibitem{bib:Graf_PhD_2008}
K. Graf, Ph.D. Thesis, Univ. Erlangen-N\"urnberg, FAU-PI1-DISS-08-001, 2008.

\bibitem{bib:graf_diplom}
K. Graf, Diploma Thesis, Univ. Erlangen-N\"urnberg, FAU-PI1-DIPL-04-002, 2004.

\bibitem{bib:Acorne2007}
Acorne Coll.,
S. Bevan \etal, 
preprint arXiv:astro-ph/0704.1025v1, 2007. 

\bibitem{bib:Thompson2008}
L. Thompson, 
% ``Acoustic detection of ultra-high-energy neutrinos''
Nucl. Instr. Meth. A {\bf 588} (2008) 155. 

\bibitem{bib:ANTARES}
ANTARES Coll., E. Aslanides \etal,
preprint arXiv:astro-ph/9907432, 1999.

\bibitem{bib:hoessl2006} G. Anton \etal, 
%``Study of piezo based sensors for acoustic particle detection'',
{\it Astropart.\ Phys.} {\bf 26} (2006) 301. 

\bibitem{bib:naumann_arena05}
C.~Naumann \etal, 
{\it Proc.\ of the Int.\ Workshop (ARENA 2005)}, 
World Scientific Publishing, Singapore (2006) 92;
preprint: arXiv:astro-ph/0511243.

\bibitem{bib:naumann_phd}
C. Naumann, Ph.D. Thesis, Univ. Erlangen-N\"urnberg, FAU-PI4-DISS-07-002, 2007.

\bibitem{bib:antares_daq} ANTARES Coll., J.A.~Aguilar \etal,
%``The data acquisition system for the ANTARES neutrino telescope'', 
{\it Nucl.\ Inst.\ Meth.} A {\bf 570}(1) (2007) 107.
%(preprint arXiv:astro-ph/0610029).

\bibitem{urick} 
R.J. Urick, {\it Principles of Underwater Sound}, Peninsula Publishing,
Los Altos, USA, 1983.


\bibitem{Neff_diplom}
M. Neff,
Diploma Thesis, Univ. Erlangen-N\"urnberg, FAU-PI1-DIPL-07-003, 2007.

\bibitem{urick2}
R.J. Urick, {\it Ambient Noise in the Sea}, Peninsula Publishing,
Los Altos, USA, 1986.

\bibitem{SAUND2008}
O. Kurahashi and G. Gratta,
Submitted to J Acoust. Soc. Am.;
preprint arXiv:astro-ph/0712.1833v1.

\bibitem{ONDE2008}
NEMO Coll., 
submitted to Deep Sea Research I;
preprint arXiv:astro-ph/0804.2913.

\bibitem{bib:ardid_these_procs}
M. Ardid, Nucl. Instr. Meth. A, these proceedings.

\bibitem{richardt_diplom}
C. Richardt, 
Diploma Thesis, Univ. Erlangen-N\"urnberg, FAU-PI4-DIPL-06-001, 2006.

% notes:

% \bibitem{label2}
% If there is a note, it should come last:
% \bibitem{label3} \note
% \end{subbibitems}

%\bibitem{}

\end{thebibliography}
\end{document}